\documentstyle[aps, psfig]{revtex}
\newcommand{\be}{\begin{equation}}
\newcommand{\ee}{\end{equation}}
\newcommand{\bea}{\begin{eqnarray}}
\newcommand{\eea}{\end{eqnarray}}
\newcommand{\bm}{\bibitem}
\newcommand{\om}{\omega}
\newcommand{\omx}{\omega_1}
\newcommand{\omy}{\omega_2}
\newcommand{\ban}{\bar{\omega}_1}
\newcommand{\al}{\alpha}
\newcommand{\bet}{\beta}
\newcommand{\lm}{\lambda}
\newcommand{\sg}{\sigma}
\newcommand{\de}{\delta}
\newcommand{\dex}{\Delta_1}
\newcommand{\dey}{\Delta_2}
\newcommand{\gz}{\gamma_0}
\newcommand{\Gm}{\Gamma}
\newcommand{\gf}{\gamma_5}
\newcommand{\ep}{\epsilon}
\newcommand{\gamu}{\gamma_{\mu}}

\newcommand{\bu}{\bar{u}}

\newcommand{\xs}{x \!\!\! /}
\newcommand{\ps}{p \!\!\! /}
\newcommand{\ks}{k \!\!\! /}
\newcommand{\kz}{k_0}
\newcommand{\pz}{p_0}
    
\newcommand{\vp}{\vec{p}} 
\newcommand{\vk}{\vec{k}}
\newcommand{\la}{\langle}
\newcommand{\ra}{\rangle}

\newcommand{\ms}{m^2}

\newcommand{\Ep}{E^{\prime}}
\newcommand{\es}{E^2}
\newcommand{\eds}{E^{\prime 2}}

\newcommand{\pr}{^\prime}
\newcommand{\sx}{S(x)}
\newcommand{\so}{S_0}
\newcommand{\sox}{S_0(x)}
\newcommand{\bd}{\bar{d}}
\newcommand{\aA}{^a_A}
\newcommand{\aAp}{^{a\pr}_{A\pr}}
\newcommand{\bB}{^b_B}
\newcommand{\bBp}{^{b\pr}_{B\pr}}
\newcommand{\cC}{^c_C}
\newcommand{\cCp}{^{c\pr}_{C\pr}}
\newcommand{\gm}{\gamma}
\newcommand{\gamn}{\gamma_{\nu}}
\newcommand{\ab}{\alpha\beta}
\newcommand{\corf}{T\,\eta_D(x)\bar\eta_{D\pr}(0)}
\newcommand{\bab}{\{a\}}
\newcommand{\bAb}{\{A\}}

\begin{document}

\title{Spectral representation and QCD sum rules for nucleon at finite  
temperature}

\author{S. Mallik\footnote{Email address: mallik@tnp.saha.ernet.in}} 
\address{Saha Institute of Nuclear Physics,
1/AF, Bidhannagar, Kolkata-700064, India} 

\author{Sourav Sarkar\footnote{Email address: sourav@veccal.ernet.in}} 
\address{Variable Energy Cyclotron Centre, 1/AF, Bidhannagar, Kolkata-700064,
 India}

%\date{} 

\maketitle

\begin{abstract} 

We examine the problem of constructing spectral representations for two
point correlation functions, needed to write down the QCD sum rules in the
medium. We suggest constructing them from the Feynman diagrams for the
correlation functions. As an example we use this procedure to write the QCD
sum rules for the nucleon current at finite temperature.
 
\end{abstract}

%\pacs{PACS number(s): }

\section{Introduction} 
\setcounter{equation}{0}

It was realized quite early in their formulation that the QCD sum rules in
medium \cite{Shap,Drukarev} have several new aspects not shared 
by the corresponding vacuum sum rules \cite{SVZ}. On the operator side, 
there appears in general new operators in
the short distance expansion of the correlation functions, due to the 
presence of the four-velocity vector of the medium \cite{Shuryak,Mallik98}. 
Even for the old
operators, the evaluation of their matrix elements in medium may need
further untested assumptions. As an example, the four-quark matrix element
in nuclear medium is evaluated by the ground state saturation, 
different from the vacuum saturation used for the 
vacuum matrix element \cite{Jin}. 
Again on the spectral side of the sum rules, unlike the vacuum
case, a communicating single particle (with medium dependent mass and
coupling) alone does not saturate the spectral function in the low energy 
region; in the medium relevant two-particle states are equally 
important. [Throughout this work we shall be
interested only in terms linear in the distribution function. For higher
order terms, contributions from multi-particle states need be included.]

The present work concerns the spectral side of the QCD sum rules.
We take a closer look at the contributions of the two-particle intermediate
state. If we draw all the 1-loop Feynman diagrams for the correlation
function, we see that in addition to the diagrams with the (single particle)
pole alone and the (two-particle) branch points alone, there are 
other (1-particle reducible) diagrams, which appear as the product of 
the pole and the branch
points. They have remainders after extracting contributions which modify the
pole parameters. It is these remainders which are
not always included in the saturation scheme, even though they 
are generally of the same order as those it retains.

We illustrate our point by discussing the nucleon QCD sum rules at finite
temperature $T$. This choice is dictated by the simplicity of 
the nucleon current correlation function at finite $T$ and the availability
of results for comparison. For $T\leq \mu$, the pion mass and zero 
nucleon chemical potential, pions dominate the heat bath. 
The leading term in the effective
chiral Lagrangian involving nucleon and pion fields describes the
interaction well in this case, allowing the calculation of shifts in the
pole position and the residue \cite{Leutwyler90}. The operator side of the 
sum rules is also simpler at finite $T$ than say, in the nuclear medium: 
To $O(T^2)$ only the Lorentz
scalar operators entering the vacuum sum rules can contribute, the
contributions of the new, non-scalar operators being at least of $O(T^3)$.
Further, the thermal average of the operators can all be reduced to their
vacuum expectation values, so that no further assumptions are required to
evaluate them, other than those needed in the vacuum case. 
What is more, these sum rules have already been
worked out \cite{Koike} and it would be interesting to see if the 
above-mentioned remainder terms bring in any new contribution.

For increased sensitivity, we work with the so-called subtracted sum rules,
i.e. the finite $T$ sum rules from which the corresponding 
vacuum sum rules have been subtracted
out, displaying only the contributions of $O(T^2)$. The spectral
representations for the sum rules are worked out for physical masses and
couplings. Only in the evaluation of the resulting integrals, do we take the
chiral limit. For completeness we have derived in an Appendix the Wilson 
coefficients for the product of nucleon currents using a simple 
coordinate space method.  

In Sec. II we describe the construction of the spectral representation for
the correlation function. In Sec. III we evaluate these contributions in the
chiral limit. Sec IV displays the sum rules and their evaluation. Sec.V
contains our concluding remarks. Appendices A and B respectively give 
the absorptive part of 1-loop diagrams at finite $T$ and the Wilson 
coefficients in the short distance expansion of two nucleon currents,  
along with the thermal averages of the relevant operators. 

\section{Construction of spectral representation}
\setcounter{equation}{0}
\renewcommand{\theequation}{2.\arabic{equation}} 

Consider the two point correlation function 
\be
\Pi (p)=i\int d^4x e^{ipx} Tr (\rho T\eta (x) \bar{\eta} (0))
\ee
of the nucleon current $\eta (x)_{D,i}$, composed of three quark fields, 
having the quantum numbers of the nucleon, the indices $D,\,i$ referring 
to spin and isospin \cite{Ioffe1,Ioffe2}. Here 
$\rho =e^{-\bet H}/ Tre^{-\bet H}$ is the thermal density matrix with
the Hamiltonian $H$ of QCD. For the spectral
representation we do not need to spell out the quark structure 
of $\eta (x)$.

A formal spectral representation for the two point function in
$E \equiv p_0$ at fixed $\vp$ may be obtained immediately
\cite{Landau,Sarkar}. 
Evaluating the trace over a complete set of eigenstates of four-momentum 
and then again inserting the same
states between the two currents to extract the $x$-dependence, one arrives at
\be
\Pi (E, \vp)= iIm \Pi_{11} (E,\vp) +{P\over {\pi}}\int_{-\infty}^{\infty}
{d \Ep coth(\bet\Ep /2) Im\Pi_{11} (\Ep, \vp) \over {\Ep-E}},
\ee
where the resulting double sum over states within the integral may be 
converted back to the Tr(ace) to get,
\be
Im \Pi_{11} ( E, \vp)={1\over 2} \int d^4x e^{ipx} Tr(\rho 
[\eta (x), \bar{\eta}(0)])
\ee  
Here the index $11$ in $Im \Pi$ reminds us that we are working in
the real time formulation of the finite temperature field
theory, where all two point functions assume the form of a $2\otimes 2$
matrix. Thus the expression (2.1) is actually the $11$-component of the
corresponding matrix. However, no information is lost by considering the
$11$-component only, as for $p^2 (=E^2-\vp~^2)$ space-like, where one writes the
QCD sum rules, there are no other independent components. We also
note that the imaginary part of the $11$- component may be written as
$tanh(\bet E/2)$ times a function, the latter coinciding with the imaginary 
part calculated in the imaginary time formulation \cite{Weldon}.
It is thus convenient to define,
\be
Im \Pi (E,\vp)= \pi^{-1} coth(\bet E/2) Im \Pi_{11} (E, \vp).
\ee

We shall write the sum rules for $\vp =0$ and in the rest frame of the heat
bath. Then we can decompose the two point function as
\be
\Pi(E, \vp =0)=\Pi_1 (E^2) +\gz E\,\Pi_2 (E^2), 
\ee
where the scalar functions have, in the notation of Eq.(2.4), the spectral 
representations,
\be
\Pi_i (\es)=\int{d\eds Im \Pi_i (\eds)\over {\eds -\es}},\quad i=1,2.
\ee
the integral running over two branch cuts, the short cut, 
$0\leq \es \leq (m-\mu)^2$ and the unitarity cut, 
$\es\geq (m +\mu)^2$, where $m$ and $\mu$ are the masses of nucleon and pion.

Eq.(2.3) does not prove convenient to calculate the spectral function.
Instead, we wish to identify the set of relevant Feynman diagrams and
calculate the spectral function directly from them. In their work on the
propagation of a nucleon through a heat bath, Leutwyler and 
Smilga \cite{Leutwyler90}
considered the set of 1-loop Feynman diagrams for the correlation function,
correcting the vacuum amplitude, shown in Figs.1-4. From these diagrams they
calculated the amplitude in the vicinity of the nucleon pole to obtain the
shifts in the nucleon mass and the residue to $O(T^2)$. The same set of
diagrams also suffice to saturate the spectral functions to the same order
in $T$ in the low energy region. Once the spectral representations of these
diagrams are constructed, one can, of course, immediately read off the
position and residue of the nucleon pole.
 
It is simple to find the spectral representations given by these 
diagrams, except for a complication with diagrams of Fig.1(b) and 
Fig.2(a) and (b). The latter
diagrams have the structure of the ${\it product}$ of a (simple or double) pole
times branch cuts, due to $N$ and $\pi N$ intermediate states. We have 
to write these contributions as a ${\it sum}$ of the pole and the branch cuts. 
In the rest of this section we perform this separation.

Let us begin with the diagrams of Fig.1. Omitting the $\eta \bar{N}$ and 
$\bar{\eta} N$ coupling constants (to be reinstated in the next section), 
their sum is
\be
-{1\over {\gz E-m}} - {1\over {\gz E-m}} \Sigma (E) {1\over {\gz E-m}}
-\cdots,
\ee
where the dots indicate the inclusion of the series of 1-particle reducible
self-energy diagrams. The self energy matrix itself may be decomposed as 
\be
\Sigma (E) = A_1 (\es)+\gz E\, A_2 (E^2).
\ee
Given the absorptive parts, to which we turn in the next section, the 
scalar functions $A_i (E^2)$ have spectral representations like those 
for $\Pi_i$ in Eq.(2.6).

\begin{figure}
\centerline{\psfig{figure=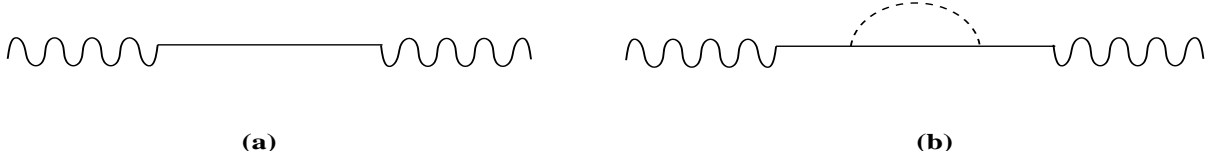,width=16cm,height=2cm}}
\caption{Nucleon pole and self-energy insertion.}
\end{figure}

\vskip 0.3in
\begin{figure}
\centerline{\psfig{figure=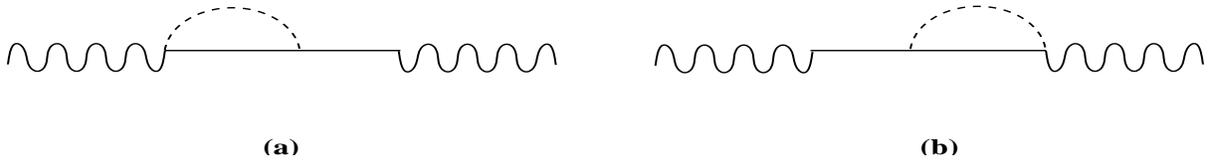,width=16cm,height=2cm}}
\caption{Vertex correction from $\pi N$ intermediate state.}
\end{figure}
 
\vskip 0.3in
\begin{figure}
\centerline{\psfig{figure=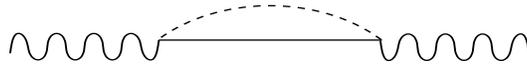,width=7cm,height=0.8cm}}
\caption{$\pi N$ intermediate state.}
\end{figure}
 
\vskip 0.3in
\begin{figure}
\centerline{\psfig{figure=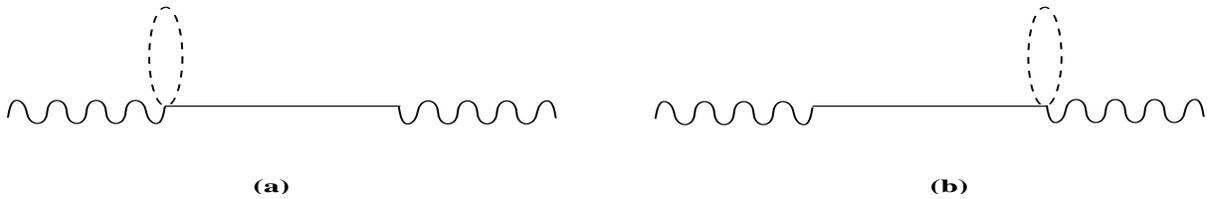,width=16cm,height=2.5cm}}
\caption{Constant vertex correction.}
\end{figure}

The second term in expression (2.7) is a product of a double pole
and cuts contained in $\Sigma$. To write it as a  sum of the pole 
and the cut contributions, we write $\Sigma (E)$ as 
\be
\Sigma (E) = A_3 (\es) +(\gz E-m) A_2 (\es), \qquad A_3 (\es)=A_1(\es) +m
A_2 (\es),
\ee
and expand the scalar functions around $\es =m^2$ to the required order
to get
\be
\Sigma (E) = a + (\es-m^2)b +(\gz E-m)c +R (E),
\ee
where, for short,
\be
 a=A_3(\ms), \ b=A_3^{\prime} (\ms), \ c=A_2(\ms),
\ee
the prime denoting differentiation with respect to $\es$. The spectral
representation for the remainder $R$ may be obtained from those of 
$A_i(\es)$ by equating the two expressions (2.9) and (2.10) for $\Sigma$. 
We get 
\be
R (E)=(\es - m^2)^2 \bar{A_3} (\es) +(\gz E -m) (\es -m^2) \bar{A_2} (\es),
\ee
where
\bea
\bar{A_3} (\es) &=&\int{d\eds Im A_3 (\eds) \over {(\eds -\ms)^2 
(\eds -\es)}}\\
\bar{A_2} (\es) &=&\int{d\eds Im A_2 (\eds) \over {(\eds -\ms) 
(\eds -\es)}} 
\eea
Inserting Eq. (2.10) in Eq.(2.7) and summing the series after isolating the
contribution from $R$, we get the desired spectral representation for the
diagrams of Fig.1 as
\be
-{1+2mb+c\over {\gz E -(m+a)}} -(\es-m^2) \bar{A_3} (\es) 
-(\gz E+m)\{2m\bar{A_3}(\es) +\bar{A_2}(\es)\}
\ee

Next consider the diagrams of Fig.2, each of which is a product of a simple
pole and cut contributions. They are of the form
\be
-{1\over {\gz E -m}}\Lambda (E),
\ee
where $\Lambda (E)$ is the vertex with $\pi N$ intermediate state. Again we
decompose $\Lambda (E)$ as
\bea
\Lambda (E)&=&B_1 (\es) +\gz E\, B_2 (\es) \nonumber \\
&=& B_3(\es)+(\gz E -m)B_2 (\es),\quad 
B_3 (\es) =B_1 (\es)+m B_2 (\es),
\eea
and expand the scalar function $B_3 (\es)$ around $\es=m^2$ to get
\be
\Lambda (E)=d+S (E),
\ee
where $d=B_3 (m^2)$ and the remainder $S$ is
\be
S(E)=(\gz E-m)\{B_2(\es) +(\gz E+m)\bar{B_3} (\es)\}
\ee
where
\be
\bar{B_3} (\es)=\int{d\eds Im B_3 (\eds) \over {(\eds -m^2) (\eds -\es)}}
\ee
The expression (2.16) now separates into the pole and the cut contributions,
\be
-{d\over {\gz E -m}} -B_2(\es) -(\gz E+m) \bar{B_3}(\es)
\ee

Clearly to find the nucleon pole position and the residue, the terms with
$A_i(E^2)$ and $B_i(E^2)$ in Eqs.(2.15, 2.21) are not necessary. But to write
the QCD sum rules, which require the correlation function for all (large, 
space-like) values of $E^2$, the entire spectral representation is needed.

\section{Evaluation of spectral representation}
\setcounter{equation}{0}
\renewcommand{\theequation}{3.\arabic{equation}}

Having separated explicitly the contributions of the $N$-pole and the $\pi
N$ cut  arising from diagrams of Fig. 1-2, we proceed to calculate the
spectral functions of all the diagrams. We need to know the 
interaction vertices present in the diagrams. The $\pi N$ interaction
Lagrangian is given, to leading order in chiral perturbation theory, by
the familiar term,
\[{\cal L}_{\pi N}={g_A\over {2F_{\pi}}} \bar{\psi} \gamma_{\mu} \tau^a \psi
\partial ^{\mu} \phi^a,\]
where $g_A=1.26$, the axial vector coupling constant of the nucleon 
and  $F_{\pi} = 93$ MeV, the pion decay constant.

The coupling of $\eta (x)$ with nucleon is defined as
\be
\la 0|\eta_i(x)|N_j(p)\ra =\de_{ij} \lm u(p) e^{-ipx}
\ee
Its couplings with nucleon and any number of pions introduce no new
constants. They may all be reduced to the coupling with nucleon alone by using
PCAC and current algebra 
\cite{Leutwyler90}. We conveniently represent these couplings by
writing $\eta (x)$ in terms of the relevant physical fields,
\be
\eta (x)=\lm \psi (x)+{i\lm\over {2F_{\pi}}} \vec{\tau}\cdot \vec{\phi}\gamma_5
\psi (x) - {\lm\over {8F_{\pi}^2}} \vec{\phi}\cdot \vec{\phi} \psi (x).
\ee

We now begin with the evaluation of the self-energy integral $\Sigma (p)$.
It is of the form of Eq. (A.1) of Appendix A with $f(p,k)$ given by 
$f(p,k)= c_1\ks\gamma_5(\ps-\ks +m)\ks \gamma_5 , \quad
c_1 = -(3 / 4) (g_A / F_{\pi})^2 $.    
Note that any term $\sim \vec{\gamma}\cdot \vk$ in $f$ will be zero
after integration over the angle of $\vk$. Then for $\vp=0$ and in the
chiral limit, it simplifies to $ f =2 c_1 k_0^2 E\gamma_0 $.
Referring to Eqs. (2.8, A.9) we get
\[Im A_1=0,\quad Im A_2= {c_1\over {4\pi^2 E}} n_1(|\ban|) |\ban|^3 ,
\qquad \ban= {E^2-m^2\over  {2E}}\]
on both the branch cuts, $ 0<\es<m^2$ and $\es >m^2$ in the chiral limit.
The constant $2m b$ is given by
\be
2mb= 2m^2\int{d\es Im A_2 (\es) \over {(\es -m^2)^2}} =-{g_A^2\over 16} 
{T^2\over {F_{\pi}^2}}
\ee
The other two constants $a$ and $c$ are each of $O(T^3)$.

The integrals in the remainder $R$ may be evaluated in the same way. Consider
$\bar{A_3}(\es)$ given by Eq.(2.13). Anticipating the QCD sum rules, we
actually need the Borel transform of the spectral representation at large
space-like values of $\es$. It is given by
\bea
2m \bar{A_3}(M^2)|_{Borel}& =& {2m^2\over M^2}\int{d\es e^{-\es /M^2} 
Im A_2 (\es) \over {(\es -\ms)^2 }},\nonumber \\
&= & - {g_A^2\over {16M^2}} e^{-m^2/M^2} {T^2\over {F_{\pi}^2}} ,
\eea
where $M$ is the Borel mass. The Borel transformed integrals for 
$(E^2-m^2) \bar{A_3} (\es)$ and $\bar{A_2}(\es)$ are each of $O(T^3)$. 

The vertex $\Lambda (E)$ is again given by Eq.(A.1) with
\[f_a (p,k)= c_2 \ks\gamma_5 (\ps-\ks +m)\gamma_5, \quad 
f_b (p,k)= c_2 \gamma_5 (\ps-\ks +m)\ks\gamma_5, \quad 
c_2 = -(3/4)g_A (\lm/F_{\pi})^2 ,\]
for diagrams (a) and (b) of Fig.2 respectively. With $\vp =0$ and chiral limit, the 
two $f$'s coincide and we have for the sum
$f (p,k)=2 c_2 k_0 (E-m\gz)$. 
Comparing with Eq.(2.17) we get
\be
Im B_{1,2} (\es) = {c_2\over {4\pi^2}} n_1(|\ban|) |\ban|\ban 
\left ( 1, -{m\over E^2} \right )
\ee
It can now be seen that both $d$ and the Borel transforms of
$B_2(\es)$ and $\bar{B}_3(\es)$ are of $O(T^4)$. [The leading pieces
from each of the two cuts are of $O(T^3)$, but they mutually cancel each
other.] Thus the diagrams of Fig.2
contribute neither to the pole nor to the cut to $O(T^2)$.

The usual contribution of the $\pi N$ intermediate state is given by 
the diagram of Fig.3,  for which $f(p,k)= c_3\gamma_5 (\ps -\ks +m)\gamma_5,
\quad c_3 = -(3/4)(\lm/F_{\pi})^2$. For $\vp =0$, the leading piece is
given by  $f = c_3 (m -\gz E)$. Then the Borel transformed amplitude is 
\be
{\lm^2\over 16} {T^2\over F_{\pi}^2}{e^{-m^2/M^2}\over M^2} (-m +\gz E).
\ee
Finally the diagram of Fig. 4 gives a constant vertex correction to the
nucleon pole,
\be
{\lm^2\over 16} {T^2\over F_{\pi}^2}{1\over {E^2 -m^2}} (m+ \gz E).
\ee

We are now in a position to write the complete spectral representation for
the two point function to $O(T^2)$. Collecting results from this and earlier
sections, the modified nucleon pole term is given by
\be
-{(\lm^T)^2\over {E^2-m^2}} (m+\gz E), 
\ee
with
\be 
\lm^T =\lm \left ( 1-{g_A^2+1\over {32}} {T^2\over {F_{\pi}^2}} \right ),
\ee
while the Borel transform of the spectral representation due to $\pi N$
branch cuts is,
\be
{\lm^2\over 16} \{ m(g_A^2-1) +(g_A^2+1)\gz E\} {T^2\over {F_{\pi}^2}}
{e^{-m^2/M^2}\over M^2}
\ee
The pole term immediately gives the results of Ref.\cite{Leutwyler90}:
the effective nucleon mass does not shift to $O(T^2)$, while the shift in 
the residue is
given by Eq.(3.9). The terms proportional to $g_A^2$ in the branch cut
contribution would have been absent, had the remainder $R$ not been taken
into account. The other remainder term $S$ does not contribute to $O(T^2)$.

\section{Sum rules}
\setcounter{equation}{0}
\renewcommand{\theequation}{4.\arabic{equation}}

The other element needed in writing down the QCD sun rules is the short
distance expansion of the product of nucleon currents. The quark content 
of the nucleon current,
most suitable for the nucleon sum rules, is given for proton (i.e. $i=1$, to
be omitted henceforth) by \cite{Ioffe2},
\be
\eta_D(x)=\ep ^{abc} (u^{a T}(x) C \gamu u^b (x) ) (\gf \gamu d^c(x))_D,
\ee
where $C$ is the charge conjugation matrix. Here $a, b, c$ are the color 
indices and $D$ is a Dirac index. 

As already stated, we shall find the finite $T$ sum rules, from which the
vacuum parts are subtracted out, thereby eliminating the contribution of the
unit operator. Since the leading thermal contribution of the gluon operators
are of $O(T^4)$ and we are working to $O(T^2)$, the contributing operators
are only $ \bu u$, $\bu \sg_{\mu\nu} t^a u G^{\mu\nu a} $ and 
$ \bu \Gm_1 u \bu\Gm_2 u$ up to dimension six. Of these the dimension five
operator has zero coefficient in the OPE of the nucleon currents
\cite{Ioffe1}. Also the thermal expectation value of the four-quark operator
turns out to be $T$-independent \cite{Koike}. Thus it is only the operator
$\bar{u} u$ which brings in any $T$-dependence on the operator side of the
sum rules. We derive systematically all these results in Appendix B. 

Usually in writing sum rules in medium, one takes the parameters of the pole
term as unknown, to be determined by the sum rules. Here these are $\lm^T$
and $m^T$, the shifted pole position.
Then the sum rules are obtained by equating the Borel
transforms of the spectral representation and the operator product
expansion. The `subtracted' sum rules read,
\bea
\overline{ (\lm^T)^2 m^T e^{-(m^T)^2/M^2}} {1\over M^2}
+{\lm^2 m\over {16M^2}} (g_A^2-1) {T^2\over {F_{\pi}^2}} e^{-m^2/M^2}
& &=\left({M\over {2\pi}}\right)^2 \la 0|\bu u|0\ra K{T^2\over
{8F_{\pi}^2}},
\nonumber \\
\overline{ (\lm^T)^2  e^{-(m^T)^2/M^2}} {1\over M^2}
+{\lm^2 \over {16M^2}} (g_A^2+1) {T^2\over {F_{\pi}^2}} e^{-m^2/M^2}
& &= 0.
\eea
where the bar over the nucleon term denotes subtraction of its vacuum value.
The factor $K$, given by
\[K=1-\left ( {W^2\over M^2}+1\right ) e^{-W^2/M^2}, \]
deviates from unity due to incorporation of the continuum contribution on
the spectral side: the imaginary part of the leading operator contribution
is assumed to saturate the spectral function from a threshold $W$ 
onwards \cite{Ioffe1}.

Let us evaluate the sum rules for $m^T$ and $\lm^T$. Write
\[ m^T = m \left (1+a {T^2\over {F_{\pi}^2}} \right ),  \quad
 \lm^T = \lm \left (1+b {T^2\over {F_{\pi}^2}} \right ) \]
where $a$ and $b$ satisfy,
\bea
b-a{m^2\over M^2} +{a\over 2}& &= -{1\over 32} (g_A^2 -1) + {1\over {16m}}
{1\over \lm^2} \left ( {M^2\over {2\pi}}\right )^2
 \la 0|\bu u | 0\ra K e^{m^2/M^2} \\
b-a{m^2\over M^2} & &= -{1\over 32} (g_A^2 +1).
\eea
The value of $\lm$ may be obtained from the vacuum sum rules. Instead of 
using the numerical estimate \cite{Ioffe1}, we may use one of the sum 
rules directly, which reads
\be
\lm^2= - {1\over {m}} \left ( {M^2\over {2\pi}}\right )^2
 \la 0|\bu u | 0\ra K e^{m^2/M^2}
\ee
Then Eq.(4.3) becomes independent of the Borel mass $M$ and one 
immediately gets,
\be
a=0, \quad b=- {1\over 32}(g_A^2+1),
\ee
reproducing the result of Ref.\cite{Leutwyler90}.

It should be noticed that the prediction (4.6) is not as clean as it
appears. Eq.(4.3) has $M$-dependence, though mild, within its allowed range
of variation. The use of the vacuum sum rule for $\lm^2$ 
compensates this variation with its own to yield the above results.

\section{Concluding remarks}
\setcounter{equation}{0}
\renewcommand{\theequation}{5.\arabic{equation}}

In this work we propose a method of constructing spectral representations
for correlation functions in the medium. It is based on its Feynman
diagrams which, to first order in the distribution function, consist
of all $1$-loop diagrams. This construction differs from the usual 
saturation scheme of taking only the single particle (with medium dependent
mass and coupling) and two-particle intermediate states by certain 
`remainder' terms contributed by the $1$-particle reducible diagrams. 

We then use this spectral representation to write down the QCD sum rules for
nucleon at finite temperature. They are shown to reproduce the temperature
dependence of the nucleon mass and its coupling to the nucleon current,
obtained earlier \cite{Leutwyler90}, justifying simultaneously the spectral
construction and the sum rules.

These sum rules were obtained earlier by Koike \cite{Koike}, though his 
analysis of the spectral function is somewhat obscured by taking the chiral 
limit from the beginning. Of the two
`remainders', he calculates one explicitly and it so turns out that the other
one, which he does not mention, is $O(T^4)$, when evaluated. As a
result his sum rules remain unaltered by our method of construction of the
spectral representation. But there are other sum rules, like the ones
for the vector mesons \cite{Hatsuda,Mallik}, which does have non-zero 
contributions from the  
`remainder' terms, not taken into account so far \cite{Mallik01}.
 
It will be observed that the Feynman diagram approach to constructing the
spectral function automatically yields also the medium dependence of the
mass and coupling of the single particle communicating with the current. It
would thus appear that the only use of the sum rules in the medium is to
rederive these results. However, the situation may not be so and one may
well be able to extract new information on the matrix elements of operators.
For example, consider the nucleon sum rules in nuclear medium. In this case 
there arises the nucleon matrix element of the four-quark operator,
mentioned already in the introduction. Usually
one relates it to the $\sigma$-term by the approximation of ground state
saturation. The sum rules, on the other hand, may give the value of this
matrix element without such an approximation.

\section*{Appendix A}
\setcounter{equation}{0}
\renewcommand{\theequation}{A.\arabic{equation}}
 
Here we derive a general formula for the discontinuity of 1-loop 
Feynman graphs encountered in this work. Each of these graphs consists of a
nucleon and a pion propagator but with different vertices. Thus they are of
the form
\be
F_{11} (p)=i\int {d^4k\over {(2\pi)^4}} f(p,k) \dex (k) \dey (p-k)
\ee
where $\Delta_{1,2}$ are the $11$-components of the corresponding propagator
matrices \cite{Kobes}; $\dex (k)$ is for pion,
\be
\dex (k) ={i\over {k^2-\mu^2 +i\ep}} +2\pi n_1(\kz) \de (k^2-\mu^2),
\ee
and $\dey (p)$ is for nucleon, after extracting the
spinor factor $(\ps + m)$,
\be
\dey (p)= {i\over {p^2-m^2 +i\ep}} -2\pi n_2(\pz)\de (p^2- m^2)
\ee
Here $n_1$ and $n_2$ are the bosonic and fermionic distribution functions:
$n_1(\kz)= (e^{\bet |\kz|}-1)^{-1}, \, n_2(\pz) = (e^{\bet |\pz|}+1)^{-1}$. 
The vertices and the spinor part of the nucleon propagator are all
contained in $f(p,k)$.

Since we work at zero chemical potential and at $T<\mu$, the nucleon
distribution function $n_2$ is negligible compared to $n_1$ for pion.
However, we wish to retain both $n_1$ and $n_2$ to show the cancellation of
their product in the final expression for the imaginary part.

We find the imaginary part of $F_{11}(p)$ by simply integrating out the time
component of $k_{\mu}$. For this purpose, we write the finite $T$ propagators 
as
\bea
\dex (k)&=& (1+n_1) D_1 (k) + n_1 D_1^*(k) \\ 
\dey (p)&=& (1-n_2) D_2 (p) -n_2 D_2^* (p)
\eea
where
\[ D_1 (k) ={i\over {k^2-\mu^2 +i\ep}},\quad 
D_2 (p) ={i\over {p^2-m^2 +i\ep}}.\]
The $k_0$ integration over the integrands $f(p,k) D_1 (k) D_2 (p-k)$ and
$f(p,k) D_1(k) D_2^* (p-k)$ are easily done by the residue theorem, getting
simple poles in $p_0$ on the real axis defined by the appropriate 
$\pm i\ep$. Discarding the principal parts, we get the imaginary part as
\bea
& &Im F_{11}(p)\nonumber \\
& & = \int {d^3k\over {(2\pi)^3}} {1\over {4\omx\omy}}\left [
\{(1+n_1)(1-n_2)-n_1n_2\}\{f(\omx) \de(E-\omx -\omy)
+f(-\omx)\de(E+\omx+\omy)\}\right.\nonumber\\
&& \left. +\{n_1(1-n_2)-(1+n_1)n_2\}
\{ f(-\omx)\de(E+\omx-\omy) + f(\omx)\de (E-\omx+\omy) \}\right]
\eea
Here
\[\omx=\sqrt{\vk^2 +\mu^2},\quad \omy =\sqrt{(\vp -\vk)^2 + m^2}. \]
and $n_1\equiv n_1(\omx)$ and $n_2 \equiv n_2(\omy)$.
For brevity, the argument of $f$ shows only the value of the integrated
variable $k_0$.

Observe the `wrong' signs in the factors involving the distribution functions,
because of which the product $n_1n_2$ does not cancel out.  It is, however,
possible to extract a factor of $tanh(\bet E/2)$ from each of the terms, by
virtue of the associated delta functions, which `corrects' the signs leading
to their cancellation. Using a notation similar to Eq.(2.4) we get
\bea
Im F(p)&=& \int {d^3k\over {(2\pi)^3}} {1\over {4\omx\omy}}\left [
(1+n_1-n_2)\{f(\omx) \de(E-\omx -\omy)
-f(-\omx)\de(E+\omx+\omy)\}\right.\nonumber\\
&&\left. +(n_1+n_2)\{f(-\omx)\de
(E+\omx-\omy) -f(\omx)\de(E-\omx+\omy)\}\right]
\eea
The positions of the branch cuts, where the imaginary 
parts are non-vanishing, are determined by the arguments of the $\de$-functions.

We next consider the limit $\vp=0$, in which we write the sum rules. The
simplified cut structure in this limit are shown in Fig.5. It is seen that
the first and the third terms in Eq.(A.7) give rise to cuts for $E>0$, and 
the second and the fourth 
terms for $E<0$. The cuts in these two regions are related by symmetry
under $E\rightarrow -E$. Restricting to cuts for $E>0$ and setting $n_2 =0$  
we get
\be
Im F(E)|_{E>0}= \int {d^3k\over {(2\pi)^3}} {1\over {4\omx\omy}}
\{(1+n_1)\,f(\omx) \de(E-\omx -\omy)
+n_1\, f(-\omx)\de(E+\omx-\omy)\}
\ee
To evaluate the integral we have, for $E>0$,
\[ {1\over {2\omy}}\de (E\mp \omx -\omy) =\de\{(E\mp\omx)^2-\omy^2\}
={1\over {2 E}} \de (\omx\mp\ban) \]
where $\ban=(E^2-m^2+\mu^2)/2E$. Note that $\ban$ is positive (negative)
on the unitary (short) cut, but the value of $\omx$ as given by the delta
functions is always positive. We thus get
\be
Im F(E)|_{E>0}= {1\over {8\pi^2 E}} g(|\ban|)\sqrt{\ban^2 -\mu^2} f(\ban),
\ee
where $g=1+n_1$ for$ E>m+\mu$ and $g=n_1$ for $0<E<m-\mu$. In writing the
sum rules, we subtract the vacuum contribution (corresponding to the term
`1' in $g$). Then $Im F(E)$ has the same expression on both the cuts.

\vskip 0.3in
\begin{figure}
\centerline{\psfig{figure=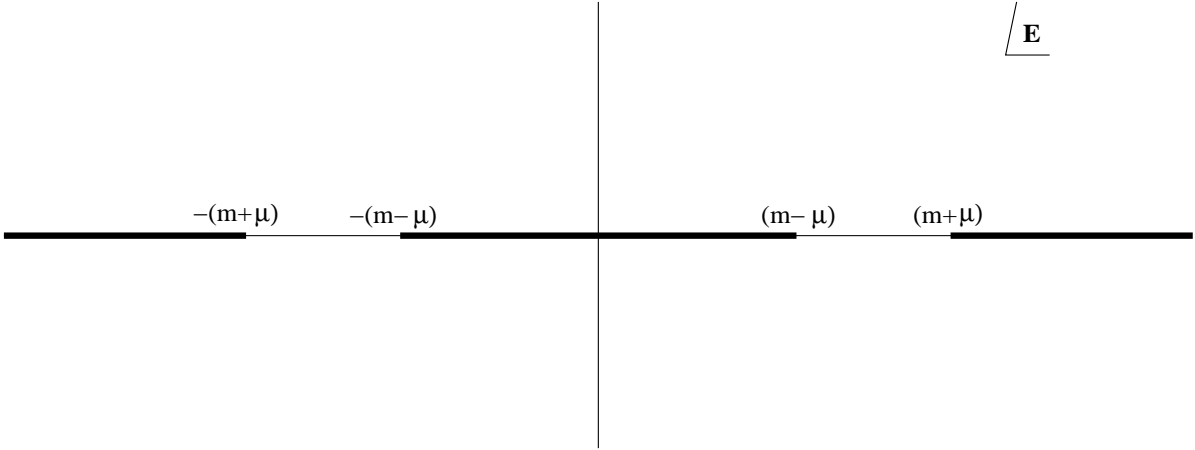,width=16cm,height=6cm}}
\caption{The cut structure in the $E$-plane}
\end{figure}
  
\section*{Appendix B} 
\setcounter{equation}{0} 
\renewcommand{\theequation}{B.\arabic{equation}}
 
Here we derive systematically the known results of operator product
expansion of nucleon currents \cite{Ioffe1}, using the co-ordinate space
method \cite{Fritzsch,Smilga,Hub,Novikov}. We also collect the vacuum and
the thermal matrix elements of the contributing operators \cite{Koike,Hatsuda}.
   
The method consists in treating the gauge field as classical and Wick  
expanding the operator product. The basic element is then the contraction of 
two, say, $u$-quark fields, 
 \[ u\overbrace{(x)\aA \bu}(0)\aAp=-i\sx^{a a\pr}_{A A\pr} \] 
where the propagator $\sx$ satisfies
\be 
-i\gm^\mu(\partial_\mu-iA_\mu(x))\sx=\delta^4(x),
\ee 
in presence of the gauge fields $A_\mu(x)=gA_\mu^n(x)\frac{\lm^n}{2}$, 
$\lm^n$'s being the Gell-Mann matrices for $SU(3)_c$. In the Fock-Schwinger 
gauge, defined by  $x^\mu A_\mu (x)=0$, the gauge potential can be expanded
in a series in the field strength $G_{\mu\nu}$ and their covariant
derivatives, 
\[A_\mu(x)=\frac{1}{2}x^\alpha G_{\al\mu}(x)+ \cdots .\] 
The crucial step in this method is to solve Eq.(B.1) in a series at short
distance \cite{Hub} 
\[\sx=\sox+S_2(x)+\cdots ,\] 
where $\sox$ is the free propagator for massless quarks, 
\[\sox^{aa\pr}_{AA\pr}=-\frac{1}{2\pi^2}\cdot\,\frac{(\xs)_{AA\pr}}{(x^2-i\ep)^2} 
\de^{aa\pr},\] 
and $S_2(x)$ is the first non-leading piece at short distance, which is 
proportional to the gauge field, 
\[S_2(x)^{aa\pr}_{AA\pr}=(\Gm^{\ab})_{AA\pr}G_{\ab}^{aa\pr},\,\,\,\,\, 
\Gm^{\ab}=-\frac{i}{16\pi^2}\cdot\,\frac{(\gm^\al\xs\gm^\bet)_{AA\pr}} 
{(x^2-i\ep)}\] 
These two terms suffice for our purpose. Note that $\sox$ is diagonal in color. 
 
With $\eta_D(x)$ given by Eq.(4.1), we have  
\[\bar\eta_{D\pr}(x)=(\bu^a(x)\gm^\nu C^{-1}\bu(x)^{bT}) 
(\bd^c(x)\gf\gm_\nu)_{D\pr}\ep^{abc}.\] 
We are interested in the two point correlation function, 
\be 
\corf=\ep^{abc}\ep^{a\pr b\pr c\pr}(C\gm^\mu)_{AB}(\gm^\nu C^{-1})_{A\pr B\pr} 
(\gf\gamu)_{DC}(\gf\gamn)_{C\pr D\pr}\,\,W^{\bab}_{\bAb}(x,0) 
\ee 
where 
\be 
W^{\bab}_{\bAb}(x,0)\equiv T u(x)^a_A u(x)^b_B d(x)^c_C \,\,\bu(0)^{a\pr}_{A\pr} 
\bu(0)^{b\pr}_{B\pr}\bd(0)^{c\pr}_{C\pr} 
\ee 
is the operator product to be expanded into local operators with 
(singular)  $c$-number coefficients as $x_\mu\rightarrow 0$. 
Its Wick expansion consists of three types of terms, 
\[W^{\bab}_{\bAb}=I^{\bab}_{\bAb}+II^{\bab}_{\bAb}+III^{\bab}_{\bAb}\] 
corresponding respectively to one, two and three contractions of the quark 
fields. 
 
The single contraction terms $I^{\bab}_{\bAb}$ may be obtained by contracting 
two $d$'s or two $u$'s. In the latter case, there arises 4 terms, all of which
are equal, as may be verified by interchanging the color indices and the Dirac 
indices in the 
terms and noting that $C\gamu$ and $\gamn C^{-1}$ are symmetric matrices. Thus 
\be 
I^{\bab}_{\bAb}(x)=-i\sx^{cc\pr}_{CC\pr}\,u(x)^a_A u(x)^b_B\bu(0)^{a\pr}_{A\pr} 
\bu(0)^{b\pr}_{B\pr} 
-4i\sx^{aa\pr}_{AA\pr}\,u(x)\bB d(x)\cC \bu(0)\bBp \bd(0)\cCp. 
\ee 
Similarly terms with two and three contractions give rise to 
\be 
II^{\bab}_{\bAb}(x)=-2\sx^{aa\pr}_{AA\pr}\sx^{bb\pr}_{BB\pr}\,\bd(0)\cCp d(x)\cC 
-4\sx^{aa\pr}_{AA\pr}\sx^{cc\pr}_{CC\pr}\,\bu(0)\bBp u(x)\bB 
\ee 
and 
\be 
III^{\bab}_{\bAb}(x)=-2i\sx^{aa\pr}_{AA\pr}\sx^{bb\pr}_{BB\pr} 
\sx^{cc\pr}_{CC\pr} 
\ee 
 
Since we do not need pure gluon operators, we do not work with Eq.(B.6). 
The quark fields in Eq.(B.4) and (B.5) may now be expanded as 
\be 
u(x)=u(0)+x^\mu D_\mu u(0) + \frac{x^\mu x^\nu}{2}D_\mu D_\nu u(0) + \cdots, 
\ee 
to get the series of local operators. Our task is now to project out the  
Lorentz scalar operators $\bu u$, $O_5\equiv\bu\sg^{\mu\nu}G_{\mu\nu}u$ and 
$\bu\Gm_1 u\bu\Gm_2u$. We consider the vacuum matrix element for this
purpose. 
 
Let us begin with $I^{\bab}_{\bAb}(x)$. As we look for operators of 
dimension no higher than six, we replace $\sx$, $u(x)$ and $d(x)$ by the  
first terms in their expansions, 
\be 
I^{\bab}_{\bAb}(x)=-iS_0(x)^{cc\pr}_{CC\pr}\,u\aA\, u\bB\,\bu\aAp\,\bu\bBp 
-4iS_0(x)^{aa\pr}_{AA\pr}\,u\bB\, d\cC \,\bu\bBp\,\bd\cCp 
\ee 
where $u\aA\equiv u\aA(0)$ etc. While projecting out the scalar part of the
4-quark operator, one simultaneously uses the approximation of vacuum
saturation to relate it to the two-quark operator. For the first term we
have \cite{SVZ}, 
\be 
\la 0|u\aA u\bB\bu\aAp\bu\bBp|0\ra=(-\de^{aa\pr}\de^{bb\pr}\de_{AA\pr} 
\de_{BB\pr}+\de^{ab\pr}\de^{ba\pr}\de_{AB\pr}\de_{BA\pr}) 
\cdot\frac{\la0|\bu u|0\ra^2}{N^2},\,\,\,\,\,\,\,\,\,\,N=12, 
\ee 
which gives rise to traces over $\gm$-matrices and sums over 
$\ep^{abc}$'s. It is simple to evaluate the traces after removing the  
$C$-matrix. We get 
\be 
\la 0|\corf|0\ra_{I,\,1st\,\,term}=\frac{i}{3\pi^2}\la0|\bu u|0\ra^2 
\cdot \frac{\xs_{DD\pr}}{(x^2)^2}. 
\ee 
A similar treatment to the second term results in a trace over an odd 
number of $\gm$-matrices, giving
\be 
\la 0|\corf|0\ra_{I,\,2nd\,\,term}=0. 
\ee

Next consider the twice contracted piece $II^{\bab}_{\bAb}$. 
The product of the  
two $\sx$'s may be expanded upto dimension two in the gauge fields to get 
\be 
\ep^{abc}\ep^{a\pr b\pr c\pr}\sx^{aa\pr}_{AA\pr}\sx^{bb\pr}_{BB\pr}= 
2\de^{cc\pr}\so\,_{AA\pr}\so\,_{BB\pr} 
-\so\,_{AA\pr}\Gm^{\ab}_{BB\pr}G_{\ab}^{c\pr c} 
-\so\,_{BB\pr}\Gm^{\ab}_{AA\pr}G_{\ab}^{c\pr c} 
\ee 
In association with the matrices $C\gm^\mu$ and$\gm^\nu C^{-1}$ in the first 
term of $II^{\bab}_{\bAb}$, the last two terms above are equal and we  
get its contribution to the operator product as, 
\bea 
\corf\mid_{II,\,1st\,\,term}&=&-4(\gf\gm^\mu)_{DC}(\gf\gm^\nu)_{C\pr D\pr} 
\left[tr(C\gamu\so\gamn C^{-1}\so^T)\left(\bd_{C\pr} d_C +  
x_\lm \bd_{C\pr}D_\lm d_C 
+\frac{x^{\lm}x^\sg}{2}\bd_{C\pr}D_\lm D_\sg d_C\right)\right.\nonumber\\ 
&&\left.-tr(C\gamu\so\gamn C^{-1}\Gm^{\ab\,\,T})\bd_{C\pr} G_{\ab}d_C 
\frac{}{}\right]. 
\eea 
Note that the color indices in the operators are now summed. 
To project out the operator $\bd d$ from Eq.(B.13), we use 
\be 
\la 0|\bd_{C\pr}d_C|0\ra=\frac{\de_{CC\pr}}{4}\la 0|\bd d|0\ra, 
\ee 
to get
\be 
\la 0|\corf |0\ra_{II,\,1st\,\,term,\,\bu u}=\frac{2}{\pi^4 x^6}\la 0|\bd d|0\ra  
\de_{DD\pr} 
\ee 
We project out the other operator $O_5$ by 
\be 
\la 0|\bd_A D_\mu D_\nu d_B|0\ra=\frac{1}{32}\left\{\de_{BA}\gm_{\mu\nu} 
-\frac{i}{3}(\sg_{\mu\nu})_{BA}\right\}\la 0|O_5|0\ra, 
\ee 
from which we also get 
\be 
\la\bd_AG_{\mu\nu}d_B\ra=\frac{1}{48}(\sg_{\mu\nu})_{BA}\la 0|O_5|0\ra . 
\ee 
It turns out that the two terms in Eq.(B.13) contributing to 
$O_5$ mutually cancel each other, 
\be 
\la 0|\corf|0\ra_{II,\,1st\,\,term,\,O_5}=0. 
\ee 
 
We are yet to find the contribution of the second term of Eq.(B.5). On using 
Eq.(B.12), it can be seen to consist of terms containing the $BB\pr$-element 
of a matrix with an odd number of $\gm$-matrices times operators 
like $\bu_{B\pr}u_B$, $\bu_{B\pr} D_\lm D_\sg u_B$ and $\bu_{B\pr}G_{\ab}u_B$. 
Also Eq.(B.14, B.16-17) show that their projection on to the
scalar operators produces a $BB\pr$-element 
of an even number of $\gm$-matrices, so we finally get a trace over an odd  
number of $\gm$-matrices. We thus get~\cite{Ioffe1} 
\be 
\la 0|\corf|0\ra_{II,\,2nd\,\,term,\,\bu u,\,O_5}=0 
\ee 
 
Thus as far as the operators $\bu u$, $O_5$ and the 4-quark 
operator are concerned, we have for the vacuum correlation function, 
\be 
i\int d^4x\,e^{ipx}\la 0|\corf|0\ra=\left(\frac{1}{2\pi}\right)^2 
p^2\ln(-p^2)\,\, 
\la 0|\bd d|0\ra \de_{DD\pr}-\frac{2}{3}\la 0|\bd d|0\ra^2\,\, 
\frac{\ps_{DD\pr}}{p^2}. 
\ee 
Since the operators concerned are Lorentz scalars,
one may think of getting the result for the thermal correlation function 
by simply replacing the vacuum expectation values in the above equation 
by their ensemble averages. However, this would not be true for the
four quark operator. For, in writing Eq.(B.9) we have already used the isospin
(and spin) structure of the vacuum matrix element, which differs from 
that for the ensemble average. A new calculation is thus necessary for
the latter.

To first order in the pion distribution function, the 
thermal average of an operator ${\cal O}$ is given by~\cite{Hatsuda} 
\be 
Tr(\rho{\cal O})=\la 0|{\cal O}|0\ra
+\sum_{i=1,2,3}\int\frac{d^3k\,n_1(\om_1)} 
{(2\pi)^3\,2\om_1}\,\,\la \pi^i(k)|{\cal O}|\pi^i(k)\ra. 
\ee 
 
On using the soft pion methods, the pion matrix element may be reduced  
to the vacuum expectation value of a double commutator, 
\be 
Tr(\rho{\cal O})=\la 0|{\cal O}|0\ra-\frac{1}{F_\pi^2} 
\int\frac{d^3k\,n_1(\om_1)}{(2\pi)^3\,2\om_1}\,\, 
\sum_{i=1,2,3}\la 0|[Q_5^i,[Q_5^i,{\cal O}]]|0\ra 
\ee 
where $Q_5^i$ is the axial-vector charge, $Q_5^i=\int d^3x A_0^i(x)$. 
An elementary way to evaluate these commutators is to write $Q_5^i$ 
in terms of quark fields, express the commutators in terms of  
anti-commutators of quark fields and replace the latter by their 
canonical values. They may then be vacuum saturated as before.
 
We record first the well-known contribution of $\bu u$ to the thermal nucleon 
correlation function, 
\be 
\Pi(p)\stackrel{\bu u}{\longrightarrow}\left(\frac{1}{2\pi}\right)^2 
\left(1-\frac{T^2}{8F_\pi^2}\right)\la 0|\bu u|0\ra\,\,p^2\,\,\ln(-p^2)\, 
\,\cdot{\bf 1} 
\ee 
For the 4-quark operators in Eq.(B.8), we have 
\bea 
&&\sum_{i=1,2,3}\la 0|\left[Q_5^i,\left[Q_5^i,u\aA u\bB u^{\dagger}\,\aAp  
u^{\dagger}\,\bBp\right]\right] |0\ra \nonumber\\ 
&&=\frac{\la0|\bu u|0\ra^2}{N^2}\cdot\left[-2\{3\gm^0_{AA\pr}\gm^0_{BB\pr} 
+(\gm^0\gf)_{AA\pr}(\gm^0\gf)_{BB\pr}\}\de^{aa\pr}\de^{bb\pr} 
-(A\pr,a\pr\,\leftrightarrow\, B\pr,b\pr)\right] 
\eea 
and 
\bea 
&&\sum_{i=1,2,3}\la 0|\left[Q_5^i,\left[Q_5^i, u\bB d\cC u^{\dagger}\,\bBp 
d^{\dagger}\,\cCp\right]\right] |0\ra \nonumber\\ 
&&=\frac{\la0|\bu u|0\ra^2}{N^2}\cdot\left[-2\{3\gm^0_{BB\pr}\gm^0_{CC\pr} 
-(\gm^0\gf)_{BB\pr}(\gm^0\gf)_{CC\pr}\}\de^{bb\pr}\de^{cc\pr} 
+4(\gm^0\gf)_{BC\pr}(\gm^0\gf)_{CB\pr}\de^{bc\pr}\de^{cb\pr}\right] 
\eea 
 
When these evaluations are inserted in the two point function, it 
turns out that the two operators bring contributions equal in magnitude but  
opposite in sign. As a result, there is no thermal contribution to $O(T^2)$ 
from the 4-quark operators, 
\be 
\Pi(p)\stackrel{4-quark}{\longrightarrow}-\frac{2}{3} 
\left(1+0\cdot\frac{T^2}{F_\pi^2}\right)\la 0|\bu u|0\ra^2\,\,\frac{\ps}{p^2} 
\ee 

\section*{Acknowledgments}

One of us (S. M.) wishes to thank H. Leutwyler for helpful 
discussions. He also acknowledges the kind hospitality at the Institute for 
Theoretical Physics, University of Berne, Switzerland.

\end{document}